\documentclass[pdflatex,iicol,sn-mathphys-num]{sn-jnl}

\usepackage{graphicx}%
\graphicspath{{./}}
\usepackage{amsmath,amssymb,amsfonts}%
\usepackage{bm}%
\usepackage{dcolumn}%
\usepackage{booktabs}%
\usepackage{textcomp}%
\usepackage{xcolor}%

\newcommand{\compactbib}{%
  \footnotesize
  \setlength{\itemsep}{-2pt}%
  \setlength{\parsep}{0pt}%
  \setlength{\parskip}{0pt}%
}

\begin{document}

\title[Effects of Mirror Dark Matter on Neutron-Star Structure and Tidal Deformability]{Effects of Mirror Dark Matter on Neutron-Star Structure and Tidal Deformability}

\author[1]{\fnm{Jin-Cheng} \sur{Jiao}}

\author*[1]{\fnm{Cheng-Ming} \sur{Li}}\email{licm@zzu.edu.cn}

\affil[1]{\orgdiv{School of Physics}, \orgname{Zhengzhou University}, \orgaddress{\city{Zhengzhou}, \postcode{450001}, \country{China}}}

\abstract{Mirror dark matter (MDM) can modify neutron-star structure and tidal response through gravitational coupling.
In this work, we construct an ordinary-matter equation of state (EOS) by comparing hadronic matter described by the relativistic mean-field NL3\(\omega\rho\) model, and quark matter in the framework of the Nambu--Jona-Lasinio (NJL) model.
The stable branch is determined through a Maxwell construction, which serves to connect distinct phases of matter.
For the parameter sets considered here, \(m_u=5.2~{\rm MeV}\) is the lowest light current-quark mass in the scanned range that satisfies the \(2M_\odot\) maximum-mass requirement, while \(m_u>5.2~{\rm MeV}\) all yield stable neutron-star configurations without a resolved macroscopic quark core.
The small-radius inferences for PSR J0437--4715 and XTE J1814--338, together with the tidal-deformability constraint from GW170817, are sensitive to the dark-matter mass fraction \(f_D\). 
The commonly used GW170817 interval \(70\lesssim\Lambda_{1.4}\lesssim580\) corresponds approximately to \(0.12\lesssim f_D\lesssim0.88\) in the present model.
These results indicate that, even without a macroscopic quark core, MDM can provide an important mechanism for reducing the visible radius and modifying the tidal response of neutron stars.}


\maketitle

\section{Introduction}

Matter inside neutron stars reaches densities far above nuclear saturation density, making these objects important astrophysical laboratories for strongly interacting matter.
At low densities, the star is mainly composed of nucleonic matter; as the density increases, hyperon degrees of freedom may appear and modify the stiffness of the hadronic equation of state (EOS).
The inclusion of hyperons usually softens the EOS and therefore affects the maximum mass and radius, so the composition of high-density hadronic matter is one of the central issues in the study of neutron-star~\cite{Oertel2017,SunZhangZhangXia2019StrangenessDelta}.
In the ordinary hadronic sector, this work considers two candidate branches: nucleonic matter and hyperonic matter.
These branches are then compared with quark matter within a unified thermodynamic construction.

From the perspective of quantum chromodynamics (QCD), the phase transition from hadronic matter to quark matter at low temperatures and high densities remains an open question.
For physical quark masses at zero baryon chemical potential, and also in the small finite-\(\mu_B\) region accessible to lattice methods, lattice QCD calculations show that observables related to chiral restoration and deconfinement vary smoothly with temperature, corresponding to a crossover rather than a genuine thermodynamic phase transition~\cite{Aoki2006,Borsanyi2020}.
At high baryon chemical potentials, however, lattice QCD is limited by the sign problem and still cannot directly determine the order of the phase transition in cold dense matter~\cite{Nagata2022}.
Many studies based on effective models demonstrate that the hadron--quark transition at zero temperature and finite chemical potentials may be either a first-order phase transition~\cite{Buballa2005,Alford2013,AlfordSedrakian2017}
or a smooth crossover~\cite{Masuda2013Crossover}.

If a first-order transition is assumed, Maxwell or Gibbs constructions are commonly used to describe thermodynamic equilibrium between the two phases~\cite{Alford2013,AlfordSedrakian2017}; if a smooth transition is assumed, crossover schemes such as three-window interpolation can be adopted~\cite{Masuda2013Crossover}.
In this work, in order to examine whether quark degrees of freedom enter stable stellar configurations, we employ the scenario of the former and use a Maxwell construction to describe the hadron--quark transition in the ordinary-matter EOS.

The properties of quark matter at high densities require an effective model description. The Nambu--Jona-Lasinio (NJL) model can describe chiral-symmetry breaking and restoration, and it is an important tool for studying quark matter at finite density~\cite{Klevansky1992,HatsudaKunihiro1994,Buballa2005,ZhangLiZong2018PNJL,LiYinZong2019PseudoWigner,ZhaoZuoLi2021NonextensivePNJL,ZhaoWangZuoLi2023NonextensivePNJL,YangFanLiMa2025CSC}.
Compared with phenomenological bag models, in the NJL framework the dynamical quark masses and the vacuum pressure term are determined by the gap equations and the vacuum structure.
This makes the model suitable for comparing the thermodynamic stability of quark matter with different flavor contents at high density.
Here, we take two-flavor and three-flavor quark matter in the NJL model into account and compare them with nucleonic matter and hyperonic matter at fixed baryon chemical potential. Then, the QCD EOS is constructed through the Maxwell construction.
In this way, the distinction among ordinary neutron stars~\cite{PangShenHu2026CrustCoreTidal,XiaXieBakhiet2024RMFConstraints},
hybrid stars~\cite{LiZhangYanHuangZong2018MassiveHybrid,ZhangGaoXiaXu2023HybridStrangeon,YuanGaoYanXu2025LargeQuarkCores,ZhangPretelXu2026SelfBoundHybrid}
and quark stars~\cite{Li2020PTR,LiZuoZhaoMuHuang2022NonstrangeQS,SuShiHuangYanLiYuanZong2024PseudoWignerQS}
is not imposed in advance but is instead determined jointly by the pressure envelope and the stellar central pressure.

Compared with studies of nonstrange two-flavor quark cores~\cite{Li2025NonstrangeCore}, the present work further includes both two-flavor and three-flavor quark matter in the EOS-construction procedure.

In addition to the phase structure of ordinary matter, dark matter may also affect the macroscopic properties of neutron stars.
Studies of dark-matter-admixed stars and two-fluid compact-star models show that dark matter can modify hydrostatic equilibrium, the radius, the maximum mass, and the tidal deformability through gravitational coupling~\cite{Leung2011,Kain2021,Leung2022,Liu2024DM}.
In mirror dark matter (MDM) models, the mirror sector has the same or similar microscopic physics as the ordinary matter sector, while the two sectors interact mainly through gravity~\cite{Foot2004,Berezhiani2004,Sandin2009MDM,Ciarcelluti2011DMCore}.
Previous studies have shown that even a small amount of MDM can significantly modify the neutron-star mass--radius relation, and that equilibrium sequences in a two-fluid system depend on the relative fraction of ordinary matter and mirror matter~\cite{Sandin2009MDM,Ciarcelluti2011DMCore,Yang2021MDMSS,Hippert2022MirrorNS,Yang2025MDMXTE}.
In this work, we model the EOS of MDM as a mirror copy of the ordinary-matter EOS and retain only gravitational coupling between the two fluids.

Recent multimessenger observations provide stringent constraints on the EOS of compact stars.
Massive pulsars such as PSR J0740+6620 require the EOS to support a maximum mass of about \(2M_\odot\)~\cite{Cromartie2020,Fonseca2021}.
The binary neutron-star merger GW170817 constrained the tidal deformability of \(1.4M_\odot\) compact stars~\cite{Abbott2017GW170817,Abbott2018Tidal,Abbott2019GW170817}.
NICER and X-ray observations provide mass--radius information for sources such as PSR J0030+0451, PSR J0740+6620, and PSR J0437--4715~\cite{Riley2019J0030,Miller2019J0030,Riley2021J0740,Miller2021J0740,Choudhury2024J0437}, while the small-radius inference for XTE J1814--338 further motivates compact configurations~\cite{Kini2024XTE,LaskosPatkos2025XTE,Yang2025MDMXTE,LopesIssifu2025}.
In two-fluid stars containing MDM, electromagnetic observations are mainly associated with the ordinary-matter radius \(R_Q\), whereas the tidal deformability refers to the total radius \(R_{\rm tot}\) and the total compactness.
Therefore, a simultaneous comparison of the \(M-R_Q\) and \(\Lambda-M\) relations between different MDM mass fractions can test the influence of MDM on neutron-star structure.

This paper is organized as follows.
In Section~II, we introduce the construction of the ordinary-matter EOS, including hadronic matter, NJL quark matter, and the Maxwell matching procedure, and also explains the treatment of the MDM EOS.
In Section~III, the two-fluid Tolman-Oppenheimer-Volkoff (TOV) equations and the calculation of the tidal Love number and dimensionless tidal deformability are presented.
In Section~IV, we provide the numerical results for the EOS of compact stars, mass--radius relations, MDM core--halo configurations, and tidal deformability.
Finally,  a summary and discussion is given in Section~V.

\section{Equations of State for Ordinary Matter and MDM}

\subsection{Ordinary QCD Matter}

The ordinary-matter EOS is constructed from hadronic matter and quark matter.
At low density, the BPS crust EOS is used to describe the matter in the outer-crust region~\cite{BaymPethickSutherland1971}; at intermediate and high densities, the hadronic matter is described by the realistic mean-field NL3\(\omega\rho\) model~\cite{Walecka1974,Lalazissis1997,HorowitzPiekarewicz2001,Oertel2017}.
We consider both NL3\(\omega\rho\) nucleonic matter without hyperons and NL3\(\omega\rho+Y\) matter including hyperon degrees of freedom.
The BPS crust is used only to complete the low-density part of the EOS and is independent of the phase-transition determination; the hyperon degrees of freedom, by contrast, is related to the phase-transition and modify the stiffness of hadronic matter at high density~\cite{Oertel2017}.

Quark matter is described by the NJL model with proper-time regularization~\cite{Nambu1961a,Nambu1961b,Klevansky1992,HatsudaKunihiro1994,Rehberg1996,Buballa2005,Wang2019PTR,Li2020PTR}.
The two-flavor quark-matter sector uses the \(SU(2)\) NJL Lagrangian
\begin{equation}
\mathcal{L}_{2f}
=
\bar q(i\gamma^\mu\partial_\mu-\hat m)q
+
G\left[
(\bar q q)^2
+
(\bar q i\gamma_5\vec{\tau}q)^2
\right],
\label{eq:NJL_Lagrangian_2f}
\end{equation}
where \(q=(u,d)^T\), \(\hat m=\mathrm{diag}(m_u,m_d)\), and \(\vec\tau\) denotes the Pauli matrices in flavor space.
In the three-flavor case, \(q=(u,d,s)^T\), and the model Lagrangian is
\begin{equation}
\begin{split}
\mathcal{L}_{3f}
=&\,\bar q(i\gamma^\mu\partial_\mu-\hat m)q
\\
&+G\sum_{a=0}^{8}
\left[
(\bar q\lambda_aq)^2
+
(\bar q i\gamma_5\lambda_aq)^2
\right]  \\
&-
K\left\{
\det_f[\bar q(1+\gamma_5)q]
+
\det_f[\bar q(1-\gamma_5)q]
\right\},
\end{split}
\label{eq:NJL_Lagrangian_3f}
\end{equation}
where \(\hat m=\mathrm{diag}(m_u,m_d,m_s)\), \(G\) is the scalar four-fermion coupling constant, and \(K\) is the coupling constant of the Kobayashi--Maskawa--'t Hooft determinant interaction.
In this work, we construct two-flavor and three-flavor quark matter separately and include both as candidate high-density phases in the Maxwell construction processes.
We take the isospin symmetry \(m_u=m_d\), and use \(m_u\) to denote the common light current-quark mass.

In the mean-field approximation, the dynamical quark mass of flavor \(i, i=u,d,s\) is determined by the stationary conditions of the thermodynamic potential:
\begin{equation}
\frac{\partial \Omega_{\rm NJL}}{\partial M_i}=0 .
\label{eq:NJL_gap}
\end{equation}
Because the NJL model is a nonrenormalizable effective theory, its momentum integrals require a regularization~\cite{Klevansky1992,Buballa2005}.
In the present calculation, the thermodynamic potential is evaluated with the proper-time regularization~\cite{Wang2019PTR,Li2020PTR}.
The same regularized thermodynamic potential determines both the dynamical quark masses and the vacuum pressure term that enters the quark-matter EOS.

For each flavor sector, the effective vacuum pressure term, just like the treatment of some previous studies~\cite{Wang2019PTR,LiYuZhaoLiZhangMaHuang2026VacuumPressureQS}, is fixed by the thermodynamic potential difference between the trivial vacuum and nontrivial vacuum.
\begin{equation}
B_{N_f}
=
\Omega_{N_f}^{\rm vac}(W)
-
\Omega_{N_f}^{\rm vac}(N),
\qquad N_f=2,3 ,
\label{eq:B_Nf}
\end{equation}
where \(W\) and \(N\) denote the Wigner solution and the Nambu solution of the quark gap equation, which is used to describe the trivial vacuum and nontrivial vacuum, respectively.
Thus, the vacuum pressure term used in this work is determined by the NJL vacuum structure rather than introduced as an external phenomenological bag parameter.

For each value of the light current-quark mass, we determine the corresponding two-flavor and three-flavor parameter sets and construct the finite-density EOSs accordingly.
Thus, changing the light current-quark mass corresponds to adopting a different EOS, rather than simply shifting a fixed EOS.

Finite-density quark matter in compact stars satisfies beta-equilibrium and electric charge neutrality.
Three-flavor quark matter satisfies
\begin{equation}
\begin{gathered}
\mu_d=\mu_s=\mu_u+\mu_e,\\
\frac{2}{3}n_u-\frac{1}{3}n_d-\frac{1}{3}n_s-n_e=0 .
\end{gathered}
\label{eq:beta_neutral_3f}
\end{equation}
For two-flavor quark matter, the strange-quark contribution is omitted, and the corresponding conditions are
\begin{equation}
\begin{gathered}
\mu_d=\mu_u+\mu_e,\\
\frac{2}{3}n_u-\frac{1}{3}n_d-n_e=0 .
\end{gathered}
\label{eq:beta_neutral_2f}
\end{equation}

To compare two-flavor and three-flavor quark matter using the same thermodynamic variable, we define the baryon chemical potential as
\begin{equation}
\mu_B=3\mu_u+2\mu_e .
\label{eq:muB_quark}
\end{equation}
For three-flavor matter, because \(\mu_d=\mu_s=\mu_u+\mu_e\), this expression is equivalent to \(\mu_B=\mu_u+\mu_d+\mu_s\).
For two-flavor matter, because \(\mu_d=\mu_u+\mu_e\), it is equivalent to the neutron-like baryon chemical potential \(\mu_B=\mu_u+2\mu_d\).
The corresponding baryon number density is
\begin{equation}
n_B=\frac{1}{3}\sum_i n_i ,
\label{eq:nB_quark}
\end{equation}
where the sum runs over the relevant quark flavors.

The quark-matter pressure is written as
\begin{equation}
P_{\rm QM}
=
-\Omega_{\rm QM}
=
\sum_i P_i+P_e-B_{\rm NJL},
\label{eq:P_QM}
\end{equation}
where \(P_e\) is the electron pressure and \(B_{\rm NJL}\) is the effective vacuum pressure term determined by the NJL vacuum structure.
In the present calculation, \(B_{\rm NJL}\) denotes \(B_{2f}\) for the two-flavor quark phase and \(B_{3f}\) for the three-flavor quark phase.
After beta-equilibrium and electric charge neutrality are imposed, the pressure of quark-matter \(P_{\rm QM}(\mu_B)\) becomes a function of the baryon chemical potential.

The energy density is obtained from the thermodynamic relation
\begin{equation}
\epsilon_{\rm QM}
=
-P_{\rm QM}
+
\sum_i\mu_i n_i
+
\mu_e n_e .
\label{eq:eps_QM}
\end{equation}
After imposing beta-equilibrium and electric charge neutrality, this expression becomes
\begin{equation}
\epsilon_{\rm QM}
=
-P_{\rm QM}
+
\mu_B n_B .
\label{eq:eps_QM_muB}
\end{equation}

Hadronic matter and quark matter are connected through a Maxwell construction.
At fixed \(\mu_B\), we compare four candidate states: nucleonic matter, hyperonic matter, two-flavor quark matter, and three-flavor quark matter.
This comparison is performed using the same thermodynamic variable, namely the grand potential density \(\Omega_\alpha(\mu_B)\) of each candidate phase.
At zero temperature, after the internal equilibrium conditions of each phase are imposed, the grand potential density and pressure satisfy \(\Omega_\alpha(\mu_B)=-P_\alpha(\mu_B)\).
Therefore, minimizing \(\Omega_\alpha(\mu_B)\) is equivalent to maximizing \(P_\alpha(\mu_B)\).
The stable branch is determined by the largest pressure:
\begin{equation}
P(\mu_B)
=
\max_{\alpha=N,\,HY,\,2f,\,3f}
P_\alpha(\mu_B).
\label{eq:Maxwell_pressure}
\end{equation}
In this sense, the Maxwell construction selects the stable EOS branch used in the stellar-structure calculation.
The corresponding energy density is taken from the same selected phase,
\begin{equation}
\epsilon(\mu_B)
=
\epsilon_{\alpha_{\rm max}}(\mu_B).
\label{eq:Maxwell_energy}
\end{equation}
In the hadronic sector, the change from the nucleonic branch to the hyperonic branch should be understood as a branch change appearing in the pressure-envelope construction, reflecting the inclusion of hyperon degrees of freedom in the adopted high-density hadronic EOS.
The high-density transition emphasized in the following is the transition from hyperonic hadronic matter to three-flavor quark matter.
In the Maxwell construction, a change of the selected branch occurs when the other candidate state becomes thermodynamically favored with increasing \(\mu_B\).
For the hadron--quark transition, the pressure is continuous at the transition point, while the energy density can be discontinuous~\cite{Alford2013,AlfordSedrakian2017}.
The final ordinary-matter EOS is then written in the form required for stellar-structure calculations, \(\epsilon=\epsilon(P)\).

\subsection{MDM}

MDM is a class of dark-matter candidates motivated by parity-symmetric extensions of the Standard Model.
In such models, the ordinary matter sector has a mirror counterpart; mirror particles have the same mass spectrum and the same form of microscopic interactions as ordinary particles, but they belong to a separate gauge sector~\cite{Foot2004,Berezhiani2004,Sandin2009MDM,Ciarcelluti2011DMCore,Yang2021MDMSS,Hippert2022MirrorNS}.
Accordingly, dense matter made of mirror baryons, mirror leptons, and the corresponding mirror strong interactions can also exist in the mirror sector.
If the mirror sector has the same microscopic physics as the ordinary sector, mirror dense matter can be described by the same EOS as ordinary QCD matter.

Based on this assumption, we treat MDM as a mirror copy of the ordinary-matter EOS:
\begin{equation}
\epsilon_D(P)=\epsilon_Q(P),
\label{eq:mirror_eos}
\end{equation}
where the subscript \(D\) denotes MDM and the subscript \(Q\) denotes the ordinary matter.
The EOS for ordinary matter is constructed in the previous subsection by employing the Maxwell construction to interface nucleonic matter, hyperonic matter, and both two- and three-flavor quark matter described by the NJL model.
Thus, the MDM component introduces no additional microscopic EOS parameters; instead, it changes the macroscopic stellar structure through its central pressure and mass fraction.

In principle, ordinary matter and mirror matter may have weak interactions beyond gravity, such as photon--mirror-photon kinetic mixing or neutron--mirror-neutron mixing.
If such interactions exist, however, their strengths are usually tightly constrained by cosmological and astrophysical observations, so their influence on the static equilibrium structure of compact stars can be treated as a higher-order effect~\cite{Foot2004,Berezhiani2004,Sandin2009MDM,Ciarcelluti2011DMCore,Yang2021MDMSS}.
Therefore, we retain only the gravitational coupling between ordinary matter and MDM and neglect possible nongravitational interactions.

In the stellar-structure calculations below, ordinary matter and MDM are treated as two interpenetrating perfect fluids that interact only through the common gravitational field.
They have their own pressures \(p_Q\), \(p_D\) and energy densities \(\epsilon_Q\), \(\epsilon_D\), and each satisfies the same EOS relation.
At the same time, both fluids contribute to the total gravitational field and jointly determine the total mass and total radius of the star.
Similar two-fluid frameworks have been used to study compact-star structures containing dark matter or MDM~\cite{Sandin2009MDM,Ciarcelluti2011DMCore,Leung2011,Kain2021,Leung2022,Hippert2022MirrorNS,Liu2024DM,Yang2021MDMSS,Yang2025MDMXTE}.

\section{Stellar Structure and Tidal Deformability with MDM}

We use natural units with \(\hbar=c=1\) and keep the gravitational constant \(G\) explicit.
Compact-star configurations containing MDM are described in a two-fluid formulation.
The two components have no direct nongravitational interaction and are coupled only through the common gravitational field.
This treatment follows the standard two-fluid description of dark-matter-admixed compact stars and MDM stars, in which the two fluids obey separate hydrostatic-equilibrium equations while sourcing the same spacetime geometry~\cite{Sandin2009MDM,Ciarcelluti2011DMCore,Leung2011,Kain2021,Yang2021MDMSS,Yang2025MDMXTE}.
The total energy density and total pressure entering the gravitational field equations are therefore
\begin{equation}
\epsilon(r)=\epsilon_Q(r)+\epsilon_D(r),
\label{eq:eps_total}
\end{equation}
\begin{equation}
p(r)=p_Q(r)+p_D(r).
\label{eq:p_total}
\end{equation}

In the two-fluid formulation, the TOV equations are written as~\cite{Tolman1939,OppenheimerVolkoff1939,Sandin2009MDM,Ciarcelluti2011DMCore,Leung2011,Kain2021,Yang2021MDMSS}
\begin{equation}
\frac{dm(r)}{dr}
=
4\pi r^2\epsilon(r),
\label{eq:tov_m}
\end{equation}
\begin{align}
\frac{dp_Q(r)}{dr}
=&-
\frac{\left[\epsilon_Q(r)+p_Q(r)\right]}
{r\left[r-2Gm(r)\right]}
\nonumber\\
&\times
\left[Gm(r)+4\pi G r^3p(r)\right],
\label{eq:tov_pq}
\end{align}
\begin{align}
\frac{dp_D(r)}{dr}
=&-
\frac{\left[\epsilon_D(r)+p_D(r)\right]}
{r\left[r-2Gm(r)\right]}
\nonumber\\
&\times
\left[Gm(r)+4\pi G r^3p(r)\right].
\label{eq:tov_pd}
\end{align}
Here, \(m(r)\) is the total gravitational mass enclosed within radius \(r\).
To calculate the MDM mass, we also introduce
\begin{equation}
\frac{dm_D(r)}{dr}
=
4\pi r^2\epsilon_D(r).
\label{eq:tov_md}
\end{equation}

For given central pressures \(p_Q(0)\) and \(p_D(0)\), these equations determine one equilibrium configuration.
Because the two central pressures can be chosen independently, a star containing MDM is no longer specified by a single central pressure, but instead belongs to a two-dimensional family of equilibrium configurations.
In the following, these configurations are organized by the dark-matter mass fraction
\[
f_D=\frac{M_D}{M}.
\]
The ordinary-matter radius and MDM radius are denoted by \(R_Q\) and \(R_D\), respectively, and the total radius is defined as
\[
R_{\rm tot}=\max(R_Q,R_D).
\]
When \(R_D<R_Q\), there is an MDM core inside the ordinary-matter component.
When \(R_D>R_Q\), the configuration has an MDM halo structure, and the outer boundary of the whole gravitational system is determined by the MDM component.
The distinction between \(R_Q\), \(R_D\), and \(R_{\rm tot}\) follows the convention used in MDM-admixed compact-star calculations, where \(R_Q\) corresponds to the observable radius of the ordinary component, while the total gravitational radius in halo configurations is set by the dark component~\cite{Yang2021MDMSS,Yang2025MDMXTE}.
The total stellar mass is \(M=m(R_{\rm tot})\), and the MDM mass is \(M_D=m_D(R_{\rm D})\).
The stellar sequences are selected using a standard turning-point criterion.
For each fixed-\(f_D\) sequence constructed from the two-dimensional central-pressure grid, the configurations are ordered by the total central density \(\rho_c=\rho_{Q,c}+\rho_{D,c}\).
The stable branch is identified as the ascending part of the sequence before the first maximum of the gravitational mass, equivalently satisfying \(dM/d\rho_c>0\).
Only this branch is retained in the following analysis, and \(M_{\max}(f_D)\) is defined as the endpoint of the corresponding stable branch.

The dimensionless tidal deformability is
\begin{equation}
\Lambda
=
\frac{2}{3}k_2 C^{-5},
\label{eq:Lambda}
\end{equation}
where the compactness is defined as
\[
C=\frac{GM}{R_{\rm tot}} .
\]
Here the radius entering the tidal compactness is the total radius \(R_{\rm tot}\), not necessarily the ordinary-matter radius \(R_Q\).
Therefore, the \(M-R_Q\) relation shown below represents the visible radius associated with ordinary matter, whereas \(\Lambda\) describes the response of the whole gravitational system to an external tidal field.
This distinction is particularly important for MDM halo configurations, for which \(R_D\) can exceed \(R_Q\).

In the two-fluid formulation, the tidal Love number \(k_2\) is written as~\cite{Hinderer2008,DamourNagar2009,BinningtonPoisson2009,Postnikov2010,Leung2022,Yang2021MDMSS}
\begin{equation}
k_2
=
\frac{8C^5}{5}
\frac{z}{F},
\label{eq:k2}
\end{equation}
with
\begin{equation}
z=(1-2C)^2\left[2-y_R+2C(y_R-1)\right],
\label{eq:z_love}
\end{equation}
\begin{equation}
\begin{aligned}
F
=&\,6C(2-y_R)+6C^2(5y_R-8)
\\
&+4C^3(13-11y_R)+4C^4(3y_R-2)
\\
&+8C^5(1+y_R)+3z\ln(1-2C).
\end{aligned}
\label{eq:F_love}
\end{equation}

Here, \(y_R\) is the tidal function evaluated at the total surface after the relevant density-jump corrections have been applied.
If the energy density just inside the total surface is nonzero and equal to \(\epsilon_s\), the surface correction is
\begin{equation}
y_R
=
y(R_{\rm tot})
-
\frac{4\pi R_{\rm tot}^3\epsilon_s}{M}.
\label{eq:y_surface}
\end{equation}
If one component terminates before the other, the same type of correction is applied at the corresponding component surface.
For example, when the MDM component terminates at \(R_D<R_Q\), the tidal function is corrected at \(R_D\) using the density discontinuity of the dark component and the total enclosed mass \(m(R_D)\).
Similarly, when the ordinary component terminates first in an MDM halo configuration, the correction is applied at \(R_Q\).
In addition, because the ordinary-matter EOS contains Maxwell transitions, the same jump condition is applied at internal energy-density discontinuities generated by the phase transition.
Thus, the tidal treatment follows the two-fluid tidal formalism used in MDM compact-star studies~\cite{Leung2022,Yang2021MDMSS}, while extending it to the present ordinary-matter EOS, which contains nucleonic matter, hyperonic matter, and NJL quark matter connected by a Maxwell construction.

The first-order equation for the tidal function is written below in an equivalent form of the standard relativistic tidal-perturbation equation used for compact stars~\cite{Hinderer2008,DamourNagar2009,BinningtonPoisson2009,Postnikov2010}.
For the two-fluid case, the source term contains the sound speeds of both components and follows the dark-matter-admixed-star treatment of Refs.~\cite{Leung2022,Yang2021MDMSS}.
The tidal function \(y(r)\) satisfies
\begin{equation}
\frac{dy(r)}{dr}
=
-\frac{y(r)^2}{r}
-
\frac{y(r)-6}{r-2Gm(r)}
-
r\mathcal{Q}(r),
\label{eq:y_eq}
\end{equation}
where
\begingroup
\small
\begin{equation}
\begin{split}
\mathcal{Q}(r)
=&\,
\frac{4\pi G r}{r-2Gm(r)}
\Bigg[
[5-y(r)]\epsilon(r)
+
[9+y(r)]p(r)
\\
&+
\frac{\epsilon_Q(r)+p_Q(r)}
{\partial p_Q(r)/\partial\epsilon_Q(r)}
\\
&+
\frac{\epsilon_D(r)+p_D(r)}
{\partial p_D(r)/\partial\epsilon_D(r)}
\Bigg]
\\
&-
4
\left[
\frac{
Gm(r)+4\pi G r^3p(r)
}{
r\left[r-2Gm(r)\right]
}
\right]^2 .
\end{split}
\label{eq:Q_love}
\end{equation}
\endgroup
Here, \(\partial p_i/\partial\epsilon_i\) is the squared sound speed of the corresponding fluid.
For given \(p_Q(0)\) and \(p_D(0)\), the tidal equation is integrated together with the two-fluid TOV equations, using the regular central boundary condition \(y(0)=2\).
This condition corresponds to the regular quadrupolar tidal perturbation near \(r=0\).
Each equilibrium configuration then yields \(M\), \(R_Q\), \(R_{\rm tot}\), \(f_D\), \(k_2\), and \(\Lambda\).

\section{Numerical Results}

This section presents the EOS construction and the stability-filtered two-fluid TOV results.
We first discuss the ordinary-matter phase structure, and then examine how MDM influence the mass--radius relation and tidal response.
In the stellar-structure results below, the displayed configurations are taken from the stable branches selected by the turning-point criterion described above.

\subsection{EOS and Phase Transitions}

\begin{figure}[!tbp]
    \centering
    \includegraphics[width=0.88\linewidth]{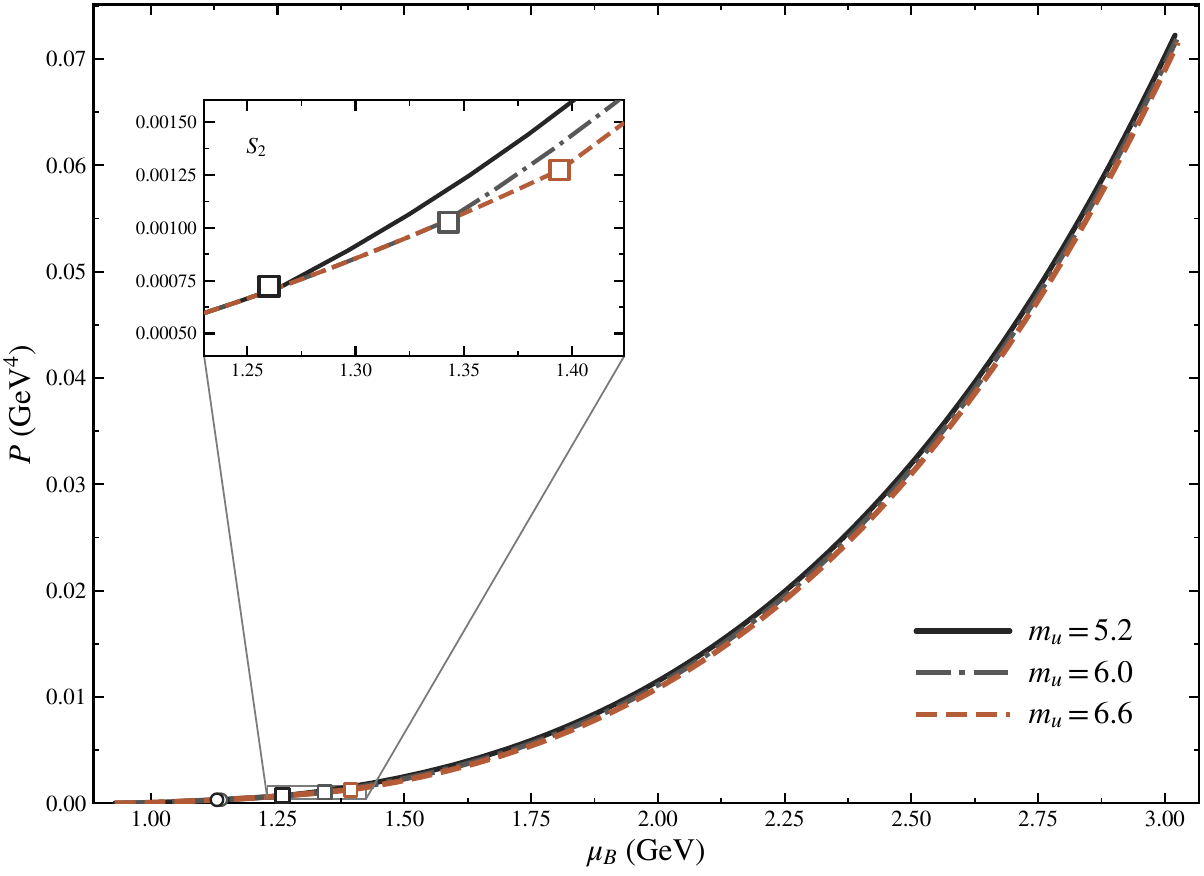}
    \caption{
    Pressure envelope \(P(\mu_B)\) obtained from the Maxwell construction for different light current-quark masses \(m_u\).
    At each given \(\mu_B\), the stable phase is selected as the branch with the largest pressure among nucleonic matter, hyperonic matter, two-flavor quark matter, and three-flavor quark matter.
    The inset magnifies the transition region \(S_2\) from hyperonic matter to three-flavor quark matter, and the open squares mark the transition points.
    }
    \label{fig:eos_muP}
\end{figure}

The first step in constructing the ordinary-matter EOS is to determine the stable branch through the Maxwell construction, rather than assuming in advance whether the star is a neutron star, a hybrid star, or a quark star.
Fig.~\ref{fig:eos_muP} shows the pressure envelope for \(m_u=5.2,6.0,6.6~{\rm MeV}\).
At each fixed value of \(\mu_B\), the stable branch is determined by the largest pressure.
The three curves nearly overlap in the low- and intermediate-chemical-potential region, indicating that the low- and intermediate-density EOSs are very similar for these representative parameters.
This is also the region that primarily determines the radii of canonical-mass neutron stars.
The main model differences are concentrated in the inset region \(S_2\), namely the transition from hyperonic matter to three-flavor quark matter.
As \(m_u\) increases, the transition point moves to higher \(\mu_B\) and higher transition pressure, so a star must reach a higher central pressure before it can encounter the quark branch.

\begin{figure}[!tbp]
    \centering
    \includegraphics[width=0.88\linewidth]{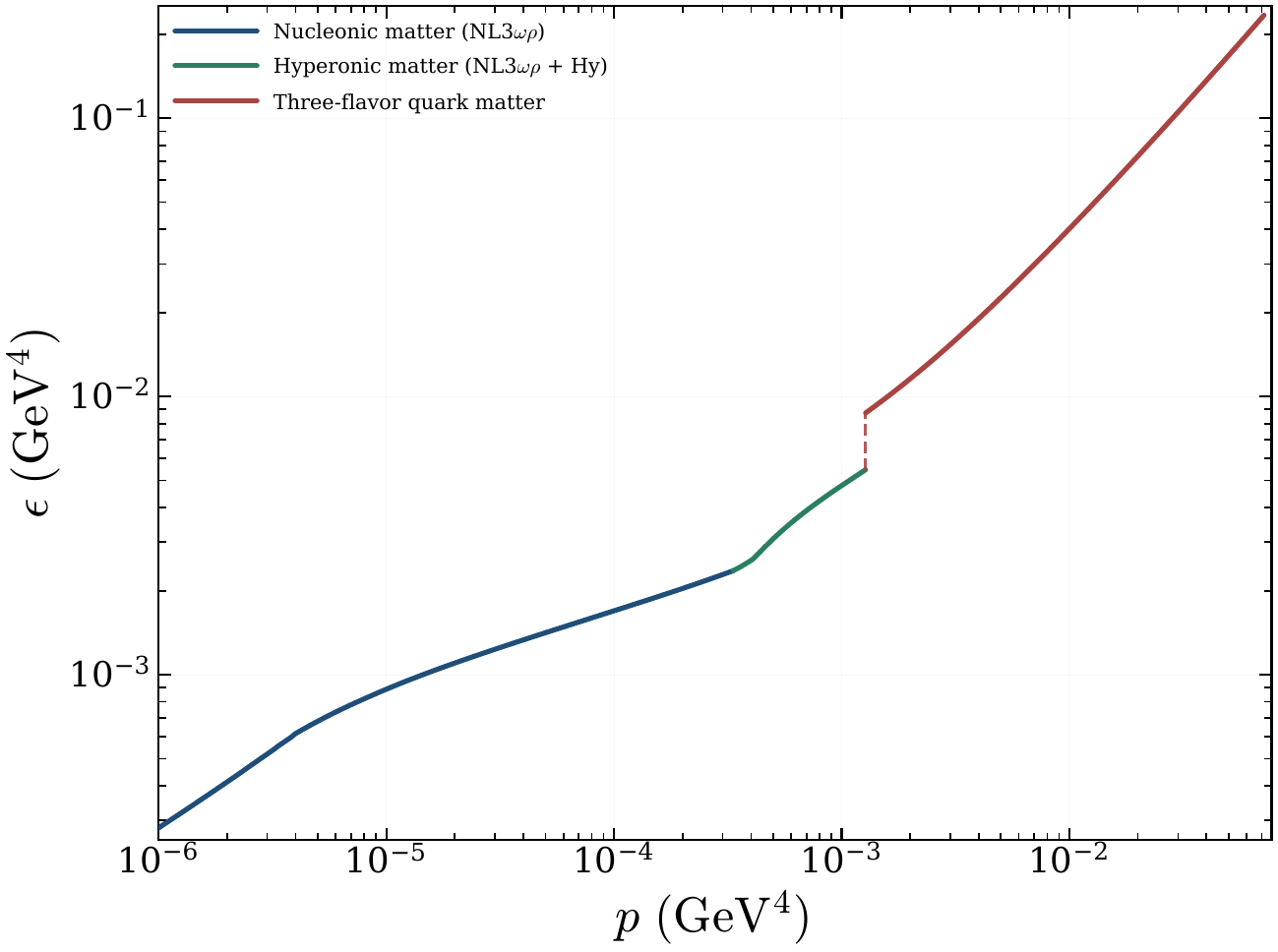}
    \caption{
    Ordinary-matter EOS after the Maxwell construction, shown in the \(\epsilon-P\) plane.
    The stable EOS branch consists of nucleonic matter, hyperonic matter, and three-flavor quark matter.
    The vertical dashed line indicates the energy-density discontinuity at the phase transition from hyperonic matter to three-flavor quark matter.
    }
    \label{fig:eos_epsP}
\end{figure}

This phase-selection result already gives an important conclusion: although two-flavor quark matter is included in the thermodynamic comparison, it does not become a stable branch of the final pressure envelope for the representative parameters adopted here.
Fig.~\ref{fig:eos_epsP} shows the same conclusion in the \(\epsilon-P\) relation.
The selected EOS passes successively through nucleonic matter, hyperonic matter, and three-flavor quark matter.
The transition from hyperonic matter to three-flavor quark matter appears as a continuous pressure with a discontinuity in energy density.
Because this discontinuity occurs at the high-pressure end, the structures of low-mass and intermediate-mass stars are mainly determined by hadronic matter, and quark degrees of freedom can affect only the central region of configurations close to the maximum mass.

\begin{figure}[!tbp]
    \centering
    \includegraphics[width=0.88\linewidth]{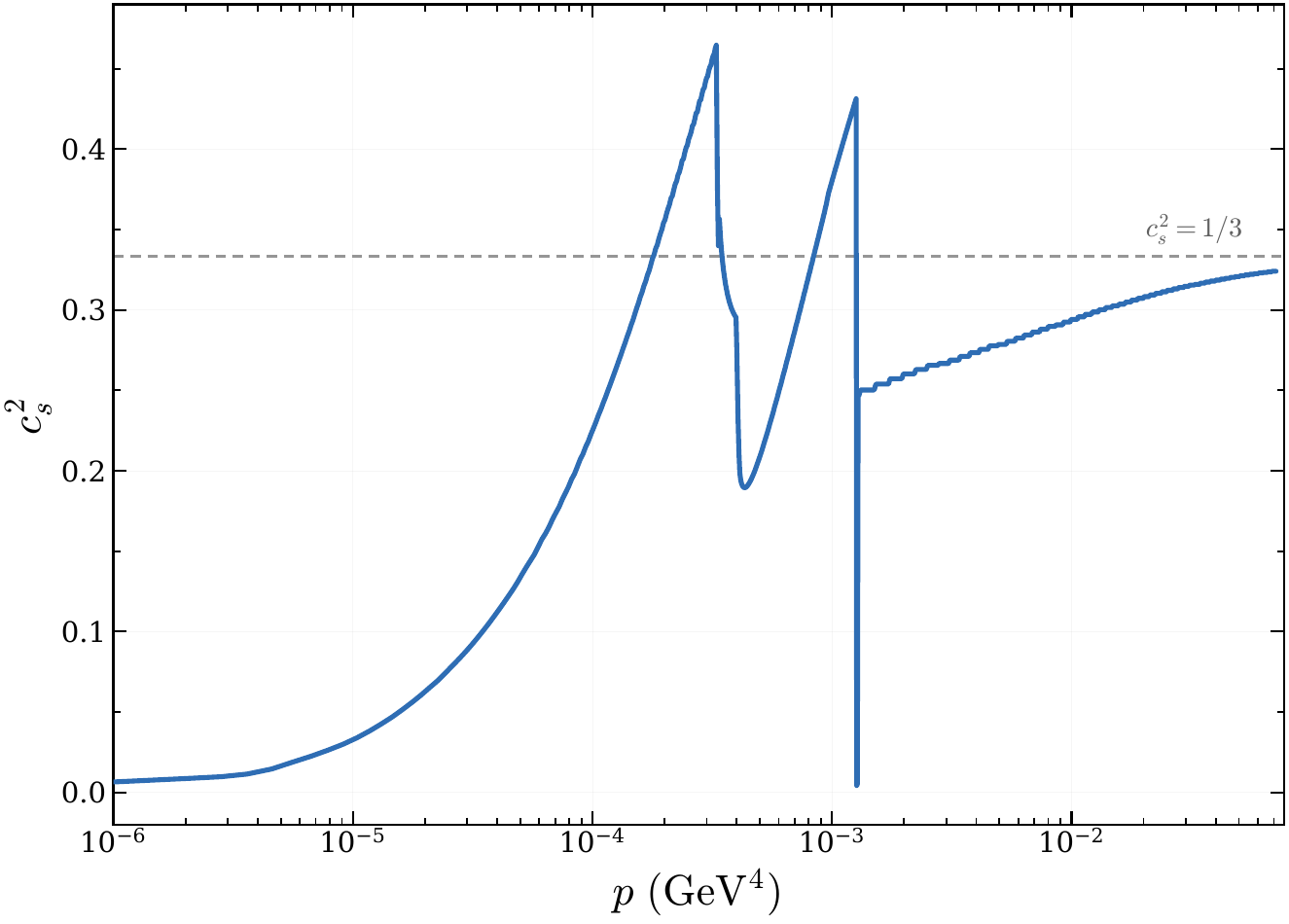}
    \caption{
    Squared sound speed \(c_s^2=dP/d\epsilon\) for the EOS of \(m_u=6.6~{\rm MeV}\) after the Maxwell construction.
    The dashed line denotes the conformal-limit reference value \(c_s^2=1/3\).
    The sharp structure at \(P\sim 10^{-4}\) GeV\(^4\) originates from the energy-density discontinuity and the change in slopes between adjacent branches.
    }
    \label{fig:eos_cs2}
\end{figure}

The squared sound speed further clarifies the pressure range over which the phase transition affects stellar structure.
The sharp structure at \(P\sim 10^{-4}\) GeV\(^4\) in Fig.~\ref{fig:eos_cs2} arises from the energy-density discontinuity and the change in slopes between adjacent branches in Fig.~\ref{fig:eos_epsP}.
At the discontinuity itself, \(dP/d\epsilon\) cannot be interpreted as the squared sound speed of a continuous medium.
At the high-pressure end, \(c_s^2\) gradually approaches the typical scale of relativistic quark matter, but this region corresponds only to configurations with high central pressure.
Thus, the behavior of \(c_s^2\) is consistent with the pressure-envelope results: the EOS contains a high-density quark phase-selection effect, but this effect does not dominate the radii of canonical-mass stars.

Table~\ref{tab:representative_mmax} lists the hadron-quark phase transition points and single-fluid maximum-mass configurations for the three representative parameter sets.
The value \(m_u=5.2~{\rm MeV}\) is the lowest light current-quark mass in the scanned range that still satisfies the maximum-mass requirement of about \(2M_\odot\), while \(m_u=6.0\) and \(6.6~{\rm MeV}\) give larger maximum masses.
None of the maximum-mass configurations in these three cases contains a resolved macroscopic quark core according to the threshold \(M_{\rm core}>10^{-3}M_\odot\).
This is consistent with Figs.~\ref{fig:eos_muP}--\ref{fig:eos_cs2}: the ordinary-matter baseline satisfying the mass constraint is not a large-scale hybrid star or a pure quark star, but rather a neutron-star-like configuration dominated by hadronic matter.

\begin{table*}[!tbp]
\centering
\caption{
Phase-transition information and maximum-mass configurations of the ordinary-matter EOSs for representative light current-quark masses.
\(P_{\rm trans}\) denotes the phase transition pressure from hyperonic matter to three-flavor quark matter;
\(M_{\max}\) is the maximum gravitational mass in the single-fluid limit;
\(R_Q(M_{\max})\) is the corresponding ordinary-matter radius.
A quark core is counted as macroscopic only when its gravitational mass exceeds \(10^{-3}M_\odot\).
}
\label{tab:representative_mmax}

\begin{tabular}{lccc}
\hline
\(m_u\) (\({\rm MeV}\)) & 5.2 & 6.0 & 6.6 \\
\hline
\(\mu_B^{\rm trans}\) (\({\rm GeV}\)) & 1.266 & 1.341 & 1.394 \\
\(P_{\rm trans}\) (\({\rm GeV}^4\)) & \(7.24\times10^{-4}\) & \(1.03\times10^{-3}\) & \(1.27\times10^{-3}\) \\
\(M_{\max}\) (\(M_\odot\)) & 2.00 & 2.08 & 2.11 \\
\(R_Q(M_{\max})\) (\({\rm km}\)) & 13.63 & 13.33 & 13.10 \\
Macroscopic quark core & No & No & No \\
\hline
\end{tabular}

\end{table*}

\subsection{Mass--Radius Relations and MDM Configurations}

\begin{figure}[!tbp]
    \centering
    \includegraphics[width=0.80\linewidth]{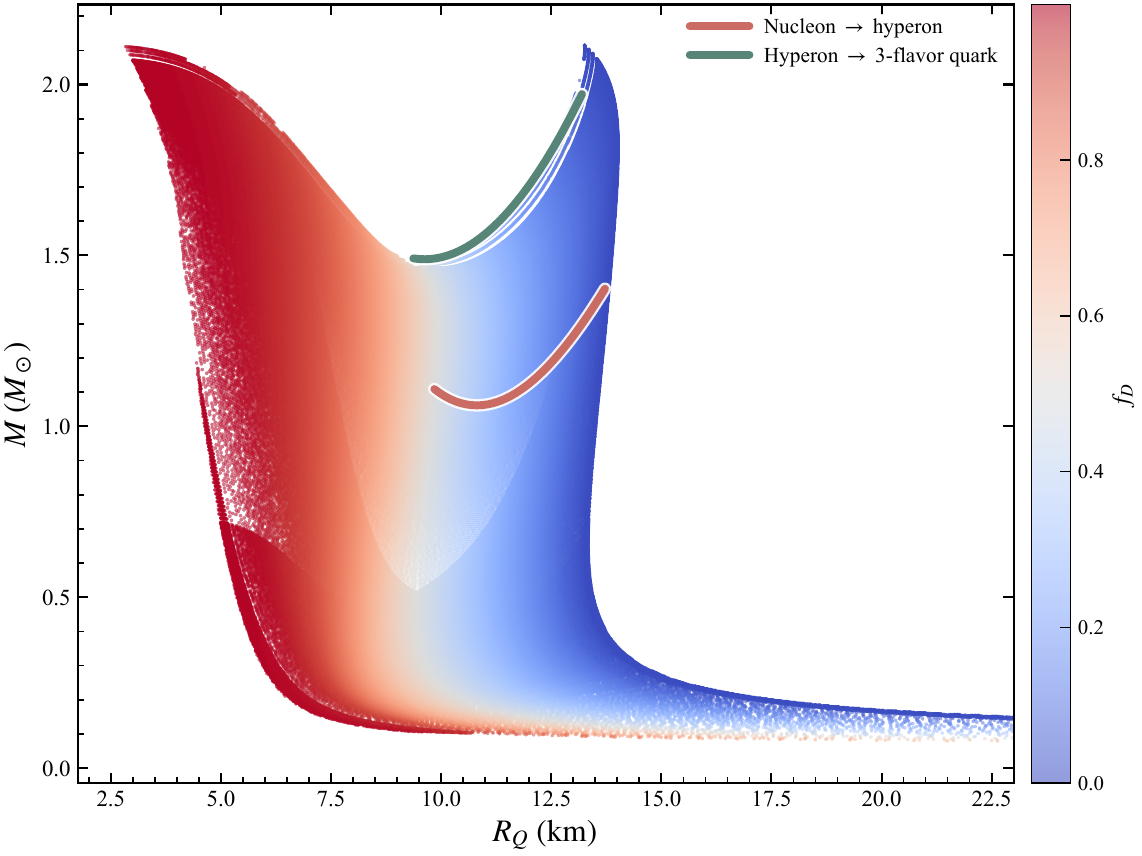}
    \caption{
    Global distribution of compact stars containing MDM in the \(M-R_Q\) plane for the EOS of \(m_u=6.6~{\rm MeV}\).
    Each scatter point corresponds to one stellar model obtained from a pair of central pressures \((p_Q(0),p_D(0))\), and the color denotes the MDM mass fraction \(f_D\).
    The solid curves show the trajectories associated with the nucleonic-to-hyperonic branch change and the hyperonic-to-three-flavor-quark transition.
    }
    \label{fig:mrq_all}
\end{figure}

In the following two-fluid calculations, unless otherwise stated, we use the ordinary-matter EOS with \(m_u=6.6~{\rm MeV}\), which gives \(M_{\max}=2.11~M_\odot\) in the single-fluid limit.
After MDM is introduced into this ordinary-matter baseline, the star is no longer uniquely determined by a single central pressure.
The central pressures of ordinary matter and mirror matter can be chosen independently, so the same EOS produces a two-dimensional distribution in the \(M-R_Q\) plane.
Fig.~\ref{fig:mrq_all} illustrates this behavior.
The plotted configurations are taken from the stable branches selected by the turning-point criterion.
The branch-change and phase-transition trajectories are mainly located at the high-mass end, whereas the color-coded \(f_D\) distribution extends over a broader mass range.
Thus, quark phase selection mainly affects the central composition near the maximum-mass region, while \(f_D\) controls how the visible component is embedded in the total gravitational field.

\begin{figure}[!tbp]
    \centering
    \includegraphics[width=0.90\linewidth]{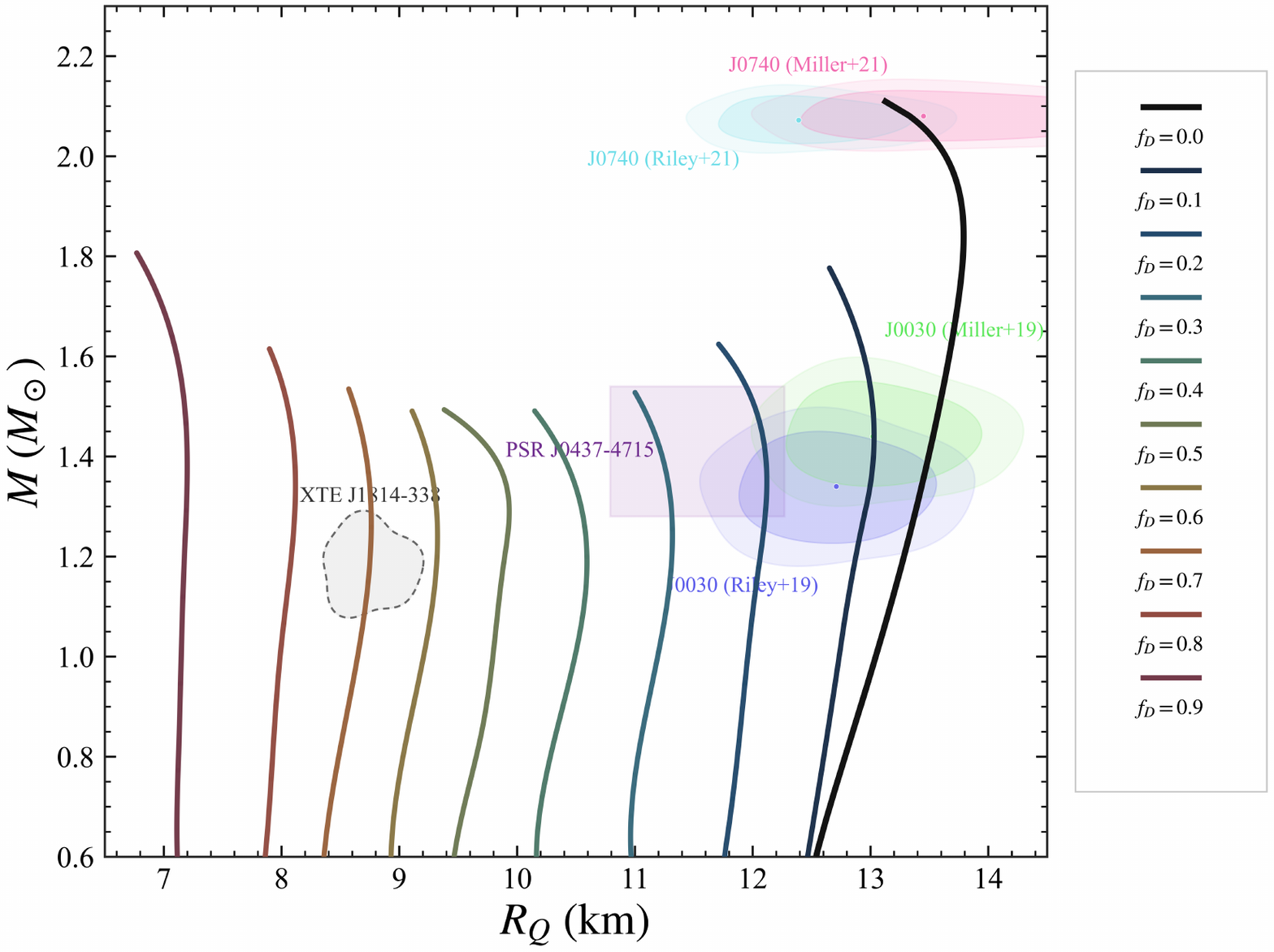}
    \caption{
    Mass--radius relations for several representative fixed values of the MDM mass fraction \(f_D\), calculated with the EOS of \(m_u=6.6~{\rm MeV}\).
    The horizontal axis \(R_Q\) denotes the ordinary-matter radius.
    The shaded regions correspond to representative mass--radius observational constraints for PSR J0030+0451, PSR J0740+6620, PSR J0437--4715, and XTE J1814--338~\cite{Riley2019J0030,Miller2019J0030,Riley2021J0740,Miller2021J0740,Choudhury2024J0437,Kini2024XTE}.
    }
    \label{fig:mrq_obs}
\end{figure}

Fig.~\ref{fig:mrq_obs} compares several fixed-\(f_D\) sequences with observational regions.
The ordinary-matter sequence with no MDM or with a low MDM fraction is already compatible with the NICER mass--radius regions of PSR J0030+0451 and PSR J0740+6620, indicating that these observations do not by themselves require a macroscopic quark core or a significant MDM component.
By contrast, the smaller-radius regions associated with PSR J0437--4715 and XTE J1814--338 are harder to reach with the ordinary-matter sequence alone.
Increasing \(f_D\) shifts the curves toward smaller \(R_Q\), allowing the model to approach these compact-radius regions while retaining massive-star support.

\begin{figure}[!tbp]
    \centering
    \includegraphics[width=0.95\linewidth]{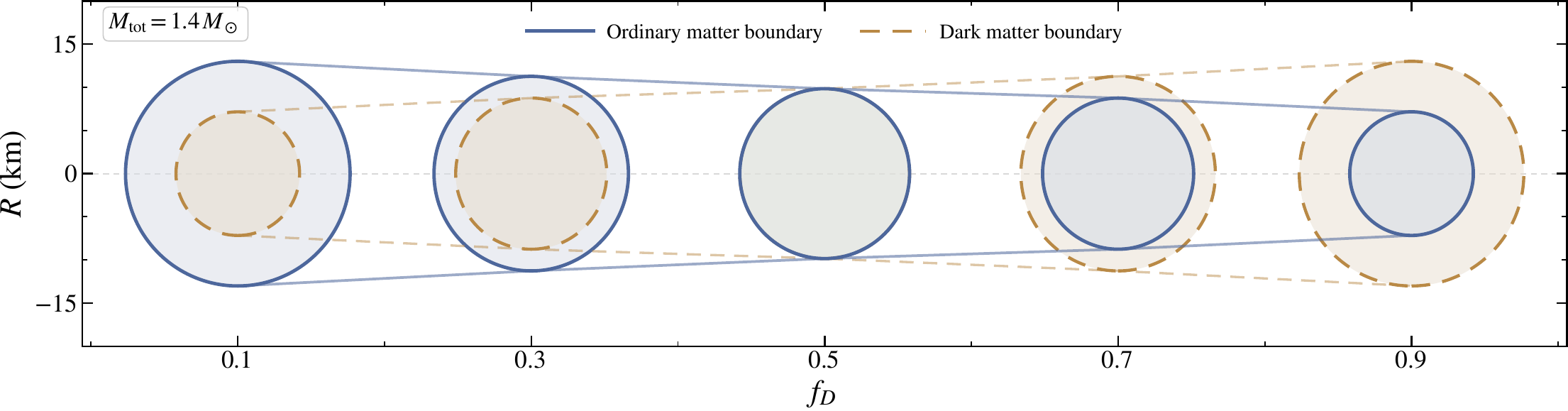}

\caption{
Schematic illustration of the ordinary-matter and MDM boundaries for configurations with fixed total mass \(M_{\rm tot}=1.4\,M_\odot\) and different MDM mass fractions \(f_D\).
As \(f_D\) increases, the configuration changes from an MDM-core structure with \(R_D<R_Q\), through the symmetric case \(R_D=R_Q\), to an ordinary-matter core surrounded by an extended MDM halo with \(R_D>R_Q\).
The connecting curves are visual guides for the evolution of the two boundaries.
Negative vertical values are used only to display the lower halves of the schematic circles and do not represent negative radii.
}

    \label{fig:mdm_config}
\end{figure}

The two-fluid configurations also require one to distinguish the ordinary-matter boundary from the total gravitational boundary.
Fig.~\ref{fig:mdm_config} schematically shows the relative locations of \(R_Q\) and \(R_D\).
When \(R_D<R_Q\), MDM mainly appears as an internal core; when \(R_D>R_Q\), MDM forms a halo configuration that extends beyond the ordinary-matter component.
The former mainly enhances the internal gravitational field and compresses the visible radius, while the latter also changes the total radius \(R_{\rm tot}\).
Therefore, the \(R_Q\) inferred from electromagnetic observations cannot be simply identified with the \(R_{\rm tot}\) entering the tidal deformability.

\subsection{Tidal Deformability}

\begin{figure}[!tbp]
    \centering
    \includegraphics[width=0.88\linewidth]{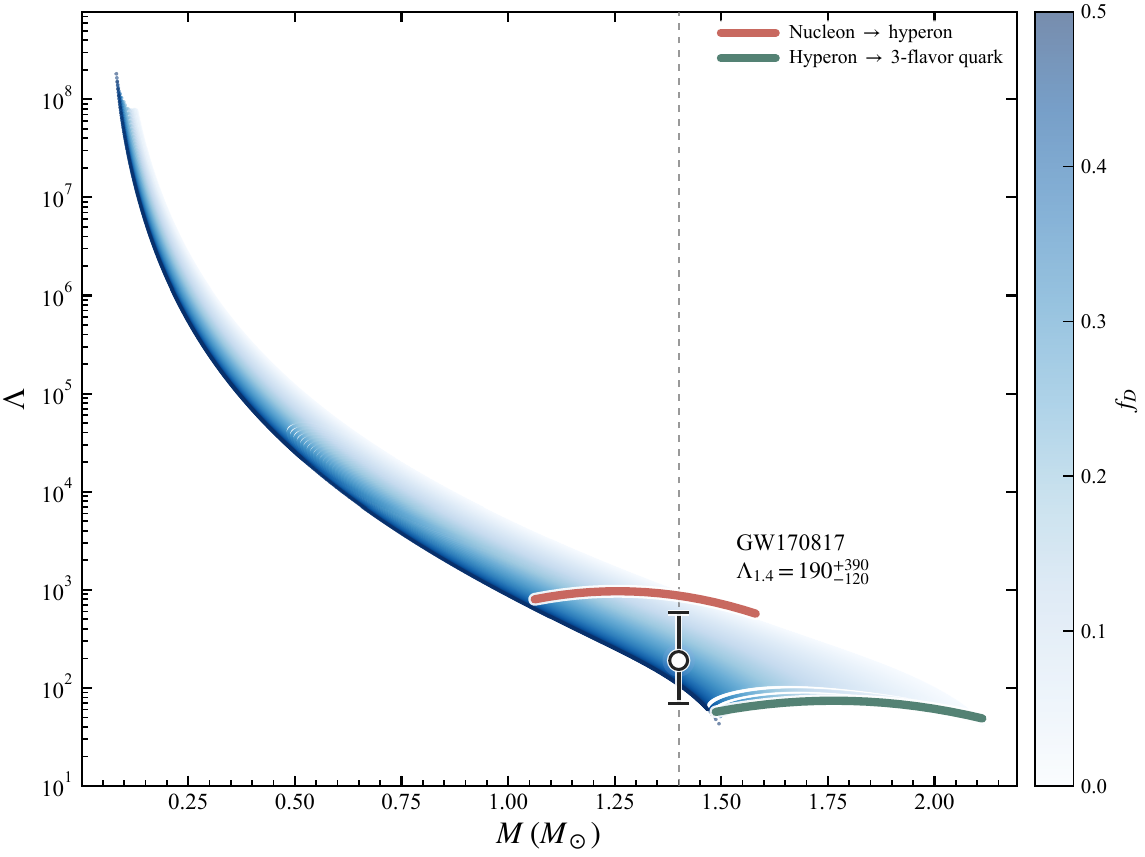}
    \caption{
    Dimensionless tidal deformability \(\Lambda\) as a function of total mass \(M\) for compact stars containing MDM, calculated with the EOS of \(m_u=6.6~{\rm MeV}\).
    The color denotes the MDM mass fraction \(f_D\).
    Since the ordinary and mirror components obey the same EOS, the tidal response is invariant under the interchange \(Q\leftrightarrow D\), corresponding to \(f_D\leftrightarrow 1-f_D\).
    The color scale therefore emphasizes the range \(0\le f_D\le0.5\), while the corresponding \(f_D>0.5\) configurations give the same \(\Lambda-M\) relation as \(1-f_D\).
    The error bar indicates the GW170817 constraint on \(\Lambda_{1.4}\)~\cite{Abbott2017GW170817,Abbott2018Tidal,Abbott2019GW170817}.
    Within the present two-fluid model, the commonly used interval \(70\lesssim\Lambda_{1.4}\lesssim580\) corresponds approximately to \(0.12\lesssim f_D\lesssim0.88\).
    The solid curves show the trajectories associated with the nucleonic-to-hyperonic branch change and the hyperonic-to-three-flavor-quark transition.
    }
    \label{fig:lambda_m}
\end{figure}

Tidal deformability probes the external gravitational response of the entire two-fluid system, rather than the ordinary-matter radius alone.
Fig.~\ref{fig:lambda_m} shows the \(\Lambda-M\) relation.
Because \(\Lambda\propto C^{-5}\), even a small change in the total compactness is strongly amplified.
For a star containing MDM, the radius entering \(C=GM/R_{\rm tot}\) is the total radius, not \(R_Q\).
Therefore, even if two models have the same mass and similar visible radii, their tidal deformabilities can differ if their MDM boundaries are different.

Fig.~\ref{fig:lambda_m} shows that \(\Lambda_{1.4}\) is relatively large near the single-fluid limit.
Two-fluid configurations in the intermediate-\(f_D\) region are more compact and therefore more likely to fall within the commonly used GW170817 interval \(70\lesssim\Lambda_{1.4}\lesssim580\).
In the present \(m_u=6.6~{\rm MeV}\) calculation, this interval maps approximately to \(0.12\lesssim f_D\lesssim0.88\).
The broad interval arises because the tidal response is symmetric under the interchange of the two identical fluids, so that configurations with \(f_D\) and \(1-f_D\) have the same total mass, total radius, Love number, and tidal deformability.
This model-dependent correspondence shows that tidal deformability is highly sensitive to the MDM mass fraction.

\subsection{Global Dependence on the MDM Fraction}

\begin{figure}[t]
    \centering
    \includegraphics[width=0.84\linewidth]{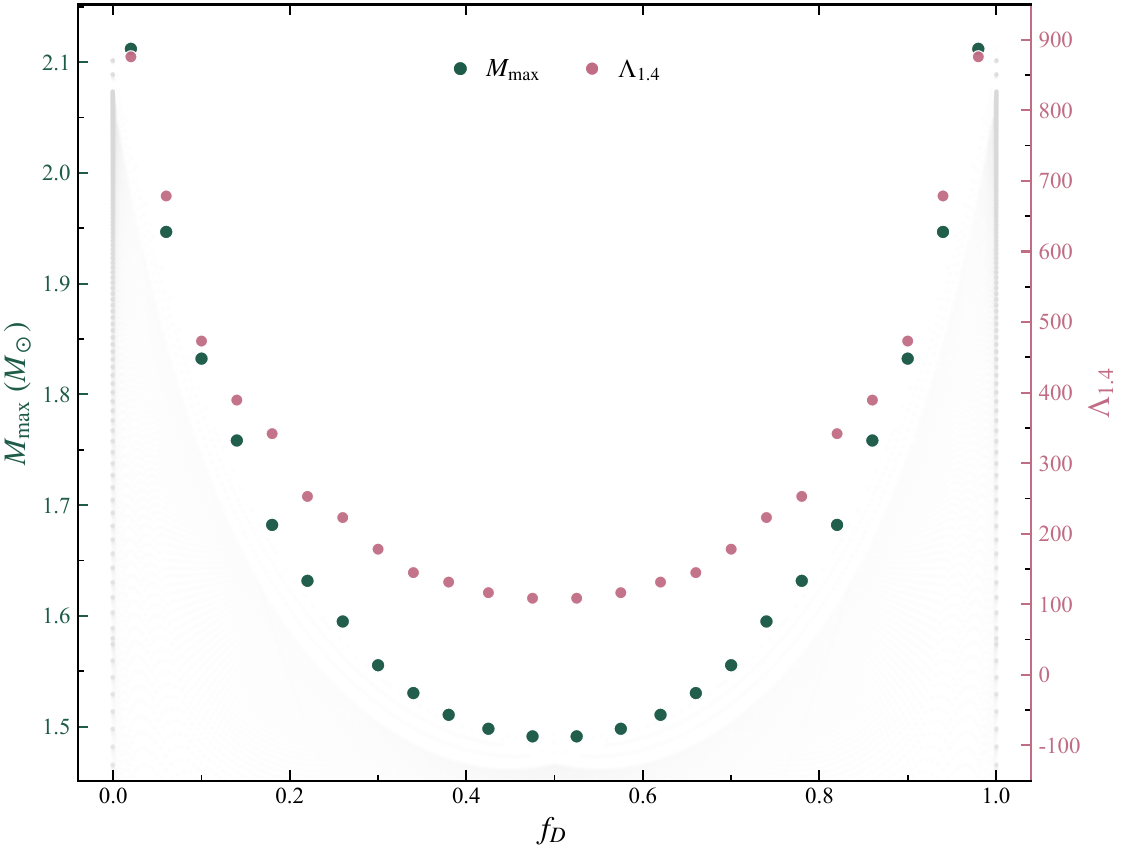}
    \caption{
    Stability-filtered maximum mass \(M_{\max}\) and dimensionless tidal deformability \(\Lambda_{1.4}\) of a \(1.4M_\odot\) star as functions of the MDM mass fraction \(f_D\), calculated with the \(m_u=6.6~{\rm MeV}\) EOS.
    The left axis corresponds to \(M_{\max}\), and the right axis corresponds to \(\Lambda_{1.4}\).
    The right vertical axis is shifted for visual separation of the two data sets; no negative tidal deformability is implied.
    }
    \label{fig:mmax_lambda14}
\end{figure}

Fig.~\ref{fig:mmax_lambda14} summarizes the dependence of the stability-filtered maximum mass and the tidal deformability of a \(1.4M_\odot\) star on \(f_D\).
Both quantities approach the single-fluid limits as \(f_D\to0\) or \(f_D\to1\), and show their largest deviations at intermediate \(f_D\), where the two components contribute comparably to the gravitational field.
This produces the nonmonotonic behavior of both \(M_{\max}\) and \(\Lambda_{1.4}\).

In the case of a single-fluid model with a hadron-quark phase transition, it is shown that the variation in the maximum mass induced by changes of the current quark mass is monotonic~\cite{Li2017Hybridstar,Li2025NonstrangeCore}.
The nonmonotonicity is the clearest two-fluid signature in the present results.
The combined behavior of \(M_{\max}\) and \(\Lambda_{1.4}\) therefore provides a useful diagnostic for separating MDM effects from a purely hadron-quark phase transition interpretation.

\section{Summary and Discussion}

This work studies the effects of mirror dark matter on neutron-star structure and tidal deformability.
The ordinary-matter EOS is obtained from a Maxwell construction among nucleonic matter, hyperonic matter, and NJL two-flavor and three-flavor quark matter.
MDM is then added as a second fluid obeying the same EOS, and only stellar configurations satisfying \(dM/d\rho_c>0\) are retained when extracting the stable branches and maximum masses.

For the representative parameter sets considered here, two-flavor quark matter does not enter the final stable pressure envelope, while three-flavor quark matter appears only at relatively high pressure.
For \(m_u=5.2,6.0,6.6~{\rm MeV}\), the single-fluid maximum masses are about \(2.00M_\odot\), \(2.08M_\odot\), and \(2.11M_\odot\), respectively.
All satisfy the massive-pulsar constraint, and none of the corresponding maximum-mass configurations contains a resolved macroscopic quark core above \(M_{\rm core}>10^{-3}M_\odot\).
Compared with studies of nonstrange two-flavor quark cores~\cite{Li2025NonstrangeCore}, this result emphasizes the importance of the quark model, flavor content, and phase-selection procedure.

In the mass--radius plane, ordinary-matter configurations can cover the NICER regions of PSR J0030+0451 and PSR J0740+6620, but they do not easily reach the smaller-radius regions associated with PSR J0437--4715 and XTE J1814--338.
Including MDM reduces the visible radius \(R_Q\) and moves the sequences toward these compact-radius regions while preserving the massive-star constraint.

Previous studies have shown that stiff hadronic EOSs may fail to satisfy the observational constraints of GW170817, and the emergence of a quark-hadron mixed phase in the core can soften the EOSs~\cite{Li2018Constraints}. In this work, we find that the presence of dark matter produces a similar effect.
Furthermore, tidal deformability gives an additional handle on \(f_D\), because \(\Lambda\) is controlled by the total compactness \(C=GM/R_{\rm tot}\), whereas electromagnetic radii mainly trace \(R_Q\).
For the present \(m_u=6.6~{\rm MeV}\) model, the symmetry between the ordinary and mirror components gives the same tidal response under \(f_D\leftrightarrow1-f_D\), and the commonly used GW170817 interval \(70\lesssim\Lambda_{1.4}\lesssim580\) maps approximately to \(0.12\lesssim f_D\lesssim0.88\).
This model-dependent mapping highlights the sensitivity of tidal observables to the MDM fraction.

The nonmonotonic dependence of \(M_{\max}\) and \(\Lambda_{1.4}\) on \(f_D\) is consistent with the two-fluid picture.
The system approaches the single-fluid limits as \(f_D\to0\) or \(f_D\to1\), while the largest deviations occur at intermediate \(f_D\), where both fluids contribute substantially to the gravitational field.

The present setup, in which ordinary matter and MDM share the same EOS and interact only through gravity, isolates the role of the MDM fraction.
Future extensions may include vector interactions in the NJL model, alternative regularization schemes, finite-size or mixed-phase effects in the phase transition, formation and capture histories for the MDM fraction, and direct comparisons with the binary tidal parameter \(\tilde\Lambda\) and Bayesian posteriors.

Overall, the massive-star constraint can be satisfied without a macroscopic quark core in the parameter range studied here, while the small-radius and tidal-deformability constraints are more sensitive to \(f_D\).
Future NICER/X-ray measurements, binary-neutron-star tidal constraints, and low-mass merger events will help distinguish a strong-phase-transition origin from the MDM two-fluid interpretation.

\backmatter

\section*{Declarations}

\noindent\textbf{Funding.}
This work is supported in part by the National Key Program for Science and Technology Research Development (2023YFB3002500), the National Natural Science Foundation of China under Grant No. 12005192, and the Natural Science Foundation of Henan Province of China (242300421375).

\noindent\textbf{Competing interests.}
The authors declare no competing interests.

\noindent\textbf{Data availability.}
All data generated or analysed during this study are included in this published article.

\noindent\textbf{Author contributions.}
Jin-Cheng Jiao is the first author and performed all calculations and wrote the manuscript. Cheng-Ming Li is the corresponding author and contributed to the research direction, supervision, and manuscript polishing suggestions.

\noindent\textbf{ORCID.}
Jin-Cheng Jiao: \href{https://orcid.org/0009-0004-4417-8596}{0009-0004-4417-8596}. Cheng-Ming Li: \href{https://orcid.org/0000-0002-9159-8129}{0000-0002-9159-8129}.

\end{document}